\documentclass[runningheads]{llncs}

\usepackage{graphicx}
\usepackage{amsmath}
\usepackage{amsfonts}
\usepackage{lipsum}
\usepackage[numbers,sort]{natbib}

\usepackage[dvipsnames]{xcolor}
\usepackage{wrapfig}

\begin{document}
\title{Counterfactual Reasoning Using Predicted Latent Personality Dimensions for Optimizing Persuasion Outcome}
\titlerunning{Counterfactual Reasoning Using Predicted Latent Personality Dimensions}
%
\author{Donghuo Zeng\inst{1}\orcidID{0000-0002-6425-6270} \and
Roberto S. Legaspi \inst{1}\orcidID{0000-0001-8909-635X} \and
Yuewen Sun \inst{2}\orcidID{0009-0002-4842-1608} \and
Xinshuai Dong \inst{2}\orcidID{0000-0002-8593-0586} \and
Kazushi Ikeda \inst{1}\orcidID{0009-0000-9563-760X} \and
Peter Spirtes \inst{2}\orcidID{0000-0002-1385-190X} \and
kun Zhang \inst{2}\orcidID{0000-0002-1343-9472}
}
\authorrunning{Donghuo Zeng et al.}
\institute{KDDI Research, Inc., Saitama, Japan \\ \email{\{do-zeng, ro-legaspi, kz-ikeda\}@kddi-research.jp} \and Carnegie Mellon University, Forbes Avenue, Pittsburgh, 15213, Pennsylvania, USA \\ \email{\{xinshuad, ps7z, kunz1\}@andrew.cmu.edu}}
\maketitle              
\begin{abstract}
Customizing persuasive conversations related to the outcome of interest for specific users achieves better persuasion results. However, existing persuasive conversation systems rely on persuasive strategies and encounter challenges in dynamically adjusting dialogues to suit the evolving states of individual users during interactions. 
This limitation restricts the system's ability to deliver flexible or dynamic conversations and achieve suboptimal persuasion outcomes. In this paper, we present a novel approach that tracks a user's latent personality dimensions (LPDs) during ongoing persuasion conversation and generates tailored counterfactual utterances based on these LPDs to optimize the overall persuasion outcome. In particular, our proposed method leverages a Bi-directional Generative Adversarial Network (BiCoGAN) in tandem with a Dialogue-based Personality Prediction Regression (DPPR) model to generate counterfactual data~$\Tilde{D}$. This enables the system to formulate alternative persuasive utterances that are more suited to the user. 
Subsequently, we utilize the D3QN model to learn policies for optimized selection of system utterances on~$\Tilde{D}$.
Experimental results we obtained from using the PersuasionForGood dataset demonstrate the superiority of our approach over the existing method, BiCoGAN. The cumulative rewards and Q-values produced by our method surpass ground truth benchmarks, showcasing the efficacy of employing counterfactual reasoning and LPDs to optimize reinforcement learning policy in online interactions.

\end{abstract}

\section{Introduction}
Persuasive conversations~\cite{prakken2006formal,torning2009persuasive,weietal2016post,yoshino2018dialogue} aim to change users' opinions, attitudes, or behaviors by employing effective communication strategies, utilizing techniques such as reasoning and emotional appeals. They involve deliberate communication with the specific goal of convincing or persuading the user to adopt a particular viewpoint or take action. In individual conversations, effective persuasion hinges on identifying unique, often hidden factors, or latent variables, such as personality (BigFive), beliefs, motivations, and experiences. Tailoring persuasive approaches to these individual nuances increases the likelihood of influencing opinions, attitudes, or behaviors.

Existing works on persuasive conversation systems have explored the impact of persuasive strategies, such as inquiry~\cite{shi2020effects,tran2022ask}, on the outcome of persuasion tasks, emphasizing strategy learning over personalized approaches. For instance, the work in~\cite{hirsh2012personalized} argues that when targeting demographics,
the sequence of persuasive strategies might be immaterial and suggests using a fixed order of persuasive appeals. However, these approaches lack flexibility and dynamics, as it relies solely on pre-defined strategy sequences while overlooking the nuanced differences in natural language. Natural language presents a challenge for persuasive conversation systems in dynamically adapting their conversations to align with individual user personalities. The complexities involved in tracking user states hinder these systems' ability to deliver adaptive and flexible dialogues.

In this work, 
we introduce an approach designed for a system to dynamically track and leverage the latent personality dimensions (LPDs) of the users during ongoing persuasive conversations. Our proposed method employs a Bi-directional Generative Adversarial Networks (BiCoGAN)~\cite{jaiswal2019bidirectional,lu2020sample} partnered with
a Dialogue-based Personality Prediction Regression (DPPR) model to generate counterfactual data $\Tilde{D}$. This enables our system to generate alternative responses fit to the user's current state, as predicted by the DPPR model. Subsequently, we optimize the system response selection using D3QN model, adapting the conversation flow to the inferred user traits.

The contribution of this work is three-fold:
(1) we trained a DPPR model to uncover the hidden aspects of their personalities over time, facilitating the tracking of user states during persuasive conversations. (2) Leveraging the BiCoGAN model with estimated individual LPDs derived from the trained DPPR model, we constructed counterfactual data~$\Tilde{D}$. This dataset provides alternative system utterances based on LPDs, extending the original dialogues of PersuasionForGood~\cite{WangSKOYZY19} dataset. (3) Employing the D3QN~\cite{raghu2017deep} model to learn policies on the counterfactual data~$\Tilde{D}$, which improves the quality of persuasive conversations, particularly in terms of enhancing persuasion outcomes.


\section{Related Work} ~\label{related}
\textbf{Persuasive strategies} identified from data gathered in persuasive online discussions and social media are frequently utilized to refine argument mining methodologies~\cite{chakrabarty2019,wachsmuth2017computational,yang2019let,wei2016post} in the construction of dialogue systems. While these works have introduced several valuable persuasion strategies, none of them have explored an efficient and automated method for applying these strategies effectively. 
Contrasting these approaches,~\cite{WangSKOYZY19} introduced the PersuasionForGood dataset through crowd-sourcing, simulating donation-related persuasive scenarios in conversational formats. \cite{shi2020effects} extended on this dataset to develop an agenda-based persuasion conversation system. However, their strategy maintained a rigid sequence of persuasive appeal strategies and lacked user modeling. To address this issue, \cite{tran2022ask} adopts reinforcement learning with dynamic user modeling to optimize the sequence of persuasive appeals based on dialogue history and the user's inclination towards donation. Persuasive systems lack fluent communication and flexibility due to strategy reliance and absence of counterfactual cases. Leveraging GANs' success in persuasive dialogue~\cite{su2018dialogue}, we generate counterfactual data to enhance adaptability.


\textbf{Personalization in persuasion}~\cite{kaptein2015personalizing,hirsh2012personalized,orji2018personalizing,rieger2022towards,matz2023potential} is important for the persuasive system to adapt to individual scenarios. Tailoring persuasive messages to align with the interests and concerns of the users is a way to enhance system effectiveness. For instance,~\cite{kaptein2015personalizing} explores the customization of persuasive technologies to individual users through persuasion profiling, utilizing both explicit measures from standardized questionnaires and implicit, behavioral measures of user traits.  
However, employing truthful personalization to enhance the persuasion dialogue may not be the optimal solution as it could impact responses over multiple steps and overlook the changing, hidden personality-related factors. Our research focuses on dynamically leveraging LPDs to capture the evolving, hidden personality-related factors during persuasive conversation. Subsequently, we employ counterfactual reasoning with LPDs to construct counterfactual data, thereby improving the persuasion outcomes.
\section{Our Architecture} ~\label{method}
\vspace{-7pt}
\begin{figure}[t]
\centering
\includegraphics[width=\textwidth]{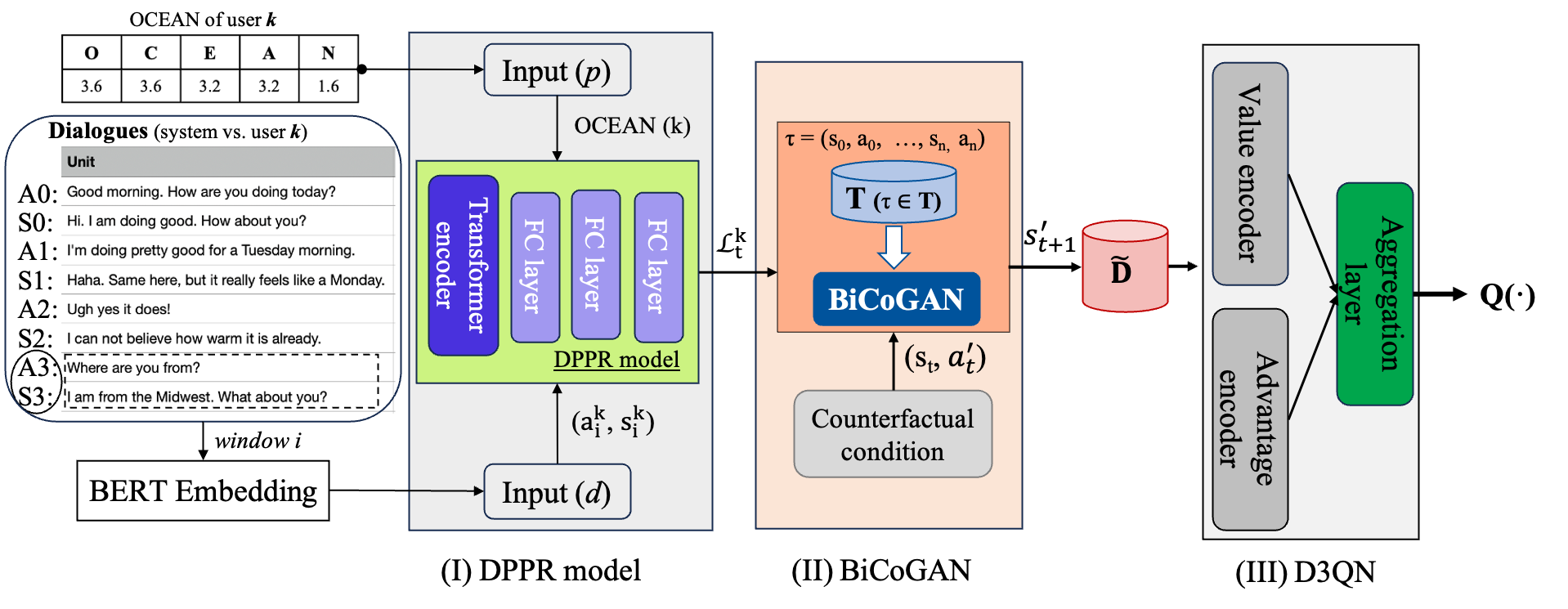}
\caption{The overview of our architecture.} \label{arch}
\vspace{-10pt}
\end{figure}

In this section, we introduce problem setting and three parts of our architecture: 1) estimation of individual latent personality dimensions, 
2) counterfactual data~$\Tilde{D}$ establishment, 
and 3) policy learning.

\subsection{Problem Setting}
Let's assume that all the dialogues are padded to the same length, and $D = \{(s_{t}, a_{t}, s_{t+1})\}^{T-1}_{t=0}$ represents an observed episode within a decision-making process governed by the dynamics of a structural causal model (SCM)~\cite{pearl2000models}, which represents causal relationships between variables via structural equations. This aims to delineate the counterfactual outcome achievable by any alternative action sequence under the specific circumstances of the episode. We explore Individualized Markov Decision Pocesses (IMDPs), in which each IMDP is defined by $M = (S, A, \mathcal{L}, f, R, T)$, where $S=\{s_{0}, s_{1}, s_{2}, ..., s_{\lceil T/2 \rceil}\}$ and $A=\{a_{0}, a_{1}, a_{2}, ..., a_{\lfloor T/2 \rfloor}\}$ are finite state and action spaces, respectively, where $\lfloor \cdot \rfloor$ represents floor function and $\lceil \cdot \rceil$ denotes ceiling function. $\mathcal{L}$ is the individual LPDs space. $R$ is the immediate reward computed by the reward model as introduced in the equation~\ref{reward_fun}. 
Accordingly, the causal mechanism $f$ is defined as
\begin{equation}
    s_{t+1} = f(s_{t}, a_{t}, \mathcal{L}_{t}, \varepsilon_{t+1})
\end{equation}
where $s_{t}$, $a_{t}$, and $\mathcal{L}_{t}$ are the state, action, and estimated LPDs at 
\begin{wrapfigure}{r}{0.5\linewidth}
\vspace{-20pt}
\includegraphics[width=\linewidth]{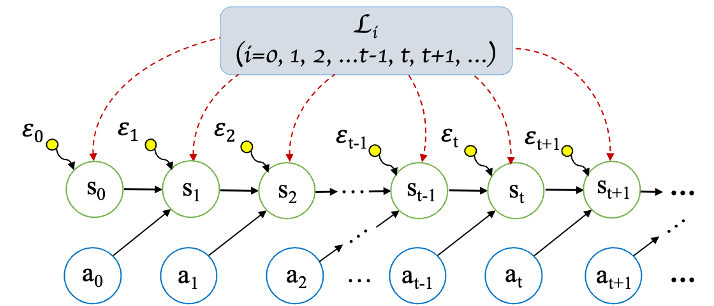}
\caption{The individualized transition dynamics model.}
\label{transition}
\vspace{-20pt}
\end{wrapfigure}
time $t$, respectively, and $\varepsilon_{t+1}$ is the noise term independent of ($s_{t}$, $a_{t}$). A graphical representation of the individual state transition process is depicted in Fig.~\ref{transition}.

Suppose $D$
is the real-world data, BiCoGAN by using estimated LPDs creates a set of $N$ counterfactual datasets $\tilde{D}=\{\tilde{D}_{0}, \tilde{D}_{1}, ..., \tilde{D}_{N}\}=\{(s^{'}_{t}, a^{'}_{t})\}^{T-1}_{t=0}$. Our target is to learn policies to optimize the persuasion outcome on counterfactual data $\tilde{D}$.

\subsection{Estimation of Individual Latent Personality Dimensions}
The OCEAN model~\footnote{The "Big Five" includes Openness, Conscientiousness, Extroversion, Agreeableness, and Neuroticism.} is widely recognized in psychology and social sciences for its reliability in assessing and understanding personality traits, which is validated across different cultures, age groups, and demographic backgrounds. However, critics argue that it falls short in representing the nuanced dynamics of personalities such as in ongoing dialogues. In addressing this, we endeavor to estimate individual LPDs in real-time to foster adaptability during conversations. It helps in dynamically adjusting persuasive approaches, such as system utterances, based on the inferred user traits for individualized conversation. For this purpose, we developed a dialogue-based personality prediction regression (DPPR) model that consists of a transformer encoder followed by three fully connected layers (refer to (I) in Fig.~\ref{arch}). During model training, the inputs comprise utterances between system and user, alongside the annotated 5-dimensional personalities OCEAN values~\cite{roccas2002big} attributed to the users. Each input is a one-turn utterance constituting a one-time exchange within a dialogue between the system and user. When a new user engages, the model progressively infers the user's LPDs over time to better understand them during the conversation. This enables the system to dynamically adapt the new utterances in the persuasion dialogue, optimizing the outcome based on the inferred individual LPDs.

\subsection{Counterfactual Data~$\tilde{D}$} \label{counter_data}
Persuasive systems heavily rely on past online interactions, often leading to sub-optimal outcomes due to the absence of sufficient actual observed results. Counterfactual reasoning enables the assessment of unobserved scenarios, encompassing various conditions and user-diverse reactions. Constructing counterfactual data facilitates enhanced policy learning, optimizing the decision-making process to achieve better outcomes.

\begin{figure}[t]
\centering
\includegraphics[width=0.9\textwidth]{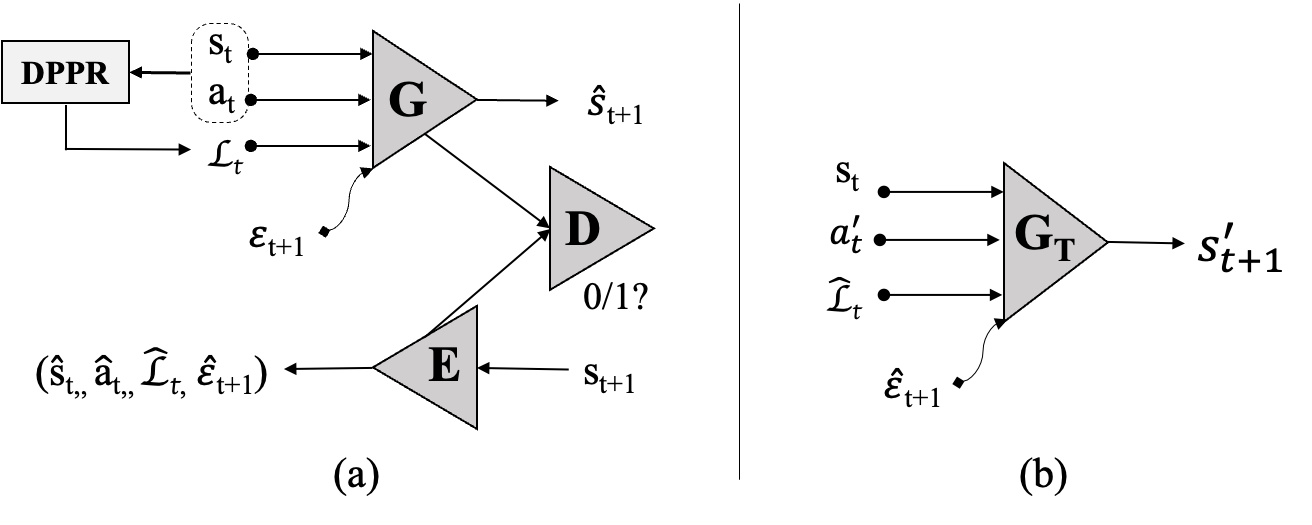}
\caption{(a) Trained DPPR model, Generator $G$, Encoder $E$, and Discriminator $D$ in the training. (b) Trained Generator $G_{T}$ in counterfactual states generating.} \label{bicogan}
\vspace{-10pt}
\end{figure}

We assume that the state $s_{t+1}$ satisfies the SCM: $s_{t+1} = f(s_{t}, a_{t}, \mathcal{L}_{t}, \varepsilon_{t+1})$, then, we employ BiCoGAN (refer to (II) in Fig.~\ref{arch})
to learn the function $f$ by minimizing the disparity between the input real data and the generated data, maintaining realistic counterfactual states aligned with observed scenarios for practical relevance. Simultaneously, it estimates the value of noise term $\varepsilon_{t+1}$, representing disturbances arising from unobserved factors, seen in Fig.~\ref{bicogan}. Specifically, the BiCoGAN proceeds in two directions: 1) mapping from ($s_{t}$, $a_{t}$, $\mathcal{L}_{t}$,  $\varepsilon_{t+1}$) to $s_{t+1}$ in the generator $G$, and 2) estimating ($s_{t}$, $a_{t}$, $\mathcal{L}_{t}$, $\varepsilon_{t+1}$) from $s_{t+1}$ via the encoder $E$. The discriminator $D$ is trained to distinguish between real and inferred data. The decoder and encoder distributions are formulated as follows, respectively.
\begin{equation}
\begin{aligned}
    P(\hat{s}_{t+1}, s_{t}, a_{t}, \mathcal{L}_{t}, \varepsilon_{t+1}) &= P(s_{t}, a_{t}, \mathcal{L}_{t}, \varepsilon_{t+1})P(\hat{s}_{t+1}|s_{t}, a_{t}, \mathcal{L}_{t}, \varepsilon_{t+1}), \\
    P(s_{t+1}, \hat{s}_{t}, \hat{a}_{t}, \hat{\mathcal{L}}_{t}, \hat{\varepsilon}_{t+1}) &= P(s_{t+1})P(\hat{s}_{t}, \hat{a}_{t}, \hat{\mathcal{L}}_{t}, \hat{\varepsilon}_{t+1}|s_{t+1})
\end{aligned}
\end{equation}
where $\hat{s}_{t}$, $\hat{s}_{t+1}$, $\hat{a}_{t}$, $\hat{\mathcal{L}}_{t}$, and $\hat{\varepsilon}_{t}$ are estimations of $s_{t}$, $s_{t+1}$, $a_{t}$, $\mathcal{L}_{t}$, and $\varepsilon_{t+1}$, respectively. The $P(\hat{s}_{t+1}|s_{t}, a_{t}, \mathcal{L}_{t}, \varepsilon_{t+1})$ and $P(\hat{s}_{t}, \hat{a}_{t}, \hat{\mathcal{L}}_{t}, \hat{\varepsilon}_{t}|s_{t+1})$ are respectively the conditional distributions of the decoder and encoder. To deceive the discriminator model, the objective function is optimized as a minimax game defined as
\begin{equation}
\begin{aligned}
\min_G \max_D V(D, G, E) &= \min_G \max_D \{\mathbb{E}_{s_{t+1} \sim p_{\text{data}}(s_{t+1})}[\log D(E(s_{t+1}), s_{t+1})] \\ &+ \mathbb{E}_{z_{t} \sim p(z_{t})}[\log(1 - D(G(z_{t}), z_{t}))] \\& +\lambda E_{(s_{t}, a_{t}, s_{t+1}) \sim p_{\text{data}}(s_{t}, a_{t}, s_{t+1})}  [R((s_{t}, a_{t}), E(s_{t+1}))]\}
\end{aligned}
\end{equation}
where $z_{t} = (s_{t}, a_{t}, \mathcal{L}_{t}, \varepsilon_{t+1})$, $R$ is a regularizer with its hyperparameter $\lambda$ to avoid overfitting issues.

After learning the SCM, counterfactual reasoning can be carried out to build the counterfactual data $\tilde{D}$. Suppose at time $t$, we have the tuple ($s_{t}$, $a_{t}$, $\mathcal{L}_{t}$, $s_{t+1}$), we want to know what would the state $s^{'}_{t+1}$ be if we take an alternative action $a^{'}_{t}$. To determine this, we take ($s_{t}$, $a^{'}_{t}$, $\mathcal{L}_{t}$), as input for the trained generator model $G_{T}$, 
which consequently outputs the counterfactual state $s^{'}_{t+1}$. Additionally, to achieve this process, we first need to build the set of counterfactual actions $\{a^{'}_{t}\}$. We derived the counterfactual actions $\{a^{'}_{t}\}$ from the real-world actions $\{a_{t}\}$ through random selection.

\subsection{Policy learning} ~\label{policy}
Following the generation of counterfactual data $\tilde{D}$, 
our approach involves learning policies on $\tilde{D}$ to maximize future rewards. We employ the Dueling Double-Deep Q-Network (D3QN)~\cite{raghu2017deep} as an enhanced variant of the standard Deep Q-Networks (DQNs) \cite{mnih2015human} to address Q-value overestimation (refer to (III) in Fig.~\ref{arch}). D3QN mitigates this issue by segregating the value function into state-dependent advantage function $A(s, a)$ and value function $V(s)$, where $A(s, a)$ capitalizes on how much better one action is than the other actions, and $V(s)$ indicates how much reward will be achieved from state $s$. The Q-values can be calculated on~$\tilde{D}$ as follows:
\begin{equation}
    \begin{aligned}
    Q(s', a';\theta) = \mathbb{E}\left[r(s', a') + \gamma \max_{a'} Q(s', a';\theta^-) \mid s^{'}, a^{'}\right],
    \end{aligned}
    \label{qlearning}
\end{equation}

\noindent where $r(s^{'}, a^{'})$ is the reward of taking action $a^{'}$ at the state $s^{'}$, the $\gamma$ is the discount factor of the max q value among all possible actions from the next state. The $r(\cdot)$ is a reward function that is used for the calculation of reward value for a given action input. The reward function is an LSTM-based architecture trained using dialogues and their corresponding donation values, the reward value of ($s'_{t}$, $a'_{t}$) is calculated as
\begin{equation}
  r(s', a')=\left\{
  \begin{array}{@{}ll@{}}
    0, & \text{if}\ t<T-1, \\
    LSTM(BERT_{embedding}(\{s'_{t}, a'_{t}\}^{T-1}_{t=0})), & \text{otherwise},
  \end{array}\right.
  \label{reward_fun}
\end{equation} 
where $T$ is the length of one dialogue in time unit. The target Q-values are derived from actions obtained through a feed-forward pass on the main network, diverging from direct estimation from the target network. During policy learning, the state transition starts with state $s^{'}_{0}$, and computing $argmax_{a^{'}_{0i}}$ $Q(s^{'}_{0}, a^{'}_{0i}; \theta, \alpha, \beta)$ leads to selecting the optimal action $a^{*}_{0}$ for state $s'_{0}$ in $\tilde{D}$, where $i=0, 1, 2, .., N-1$. In the end, we apply the Mean Squared Error (MSE) as a loss function to update the weights of D3QN neural networks.

\section{Experiment} ~\label{experiment}
\vspace{-20pt}
\subsection{Dataset}
\begin{figure}[h]
\vspace{-15pt}
\centering
\includegraphics[width=0.95\textwidth]{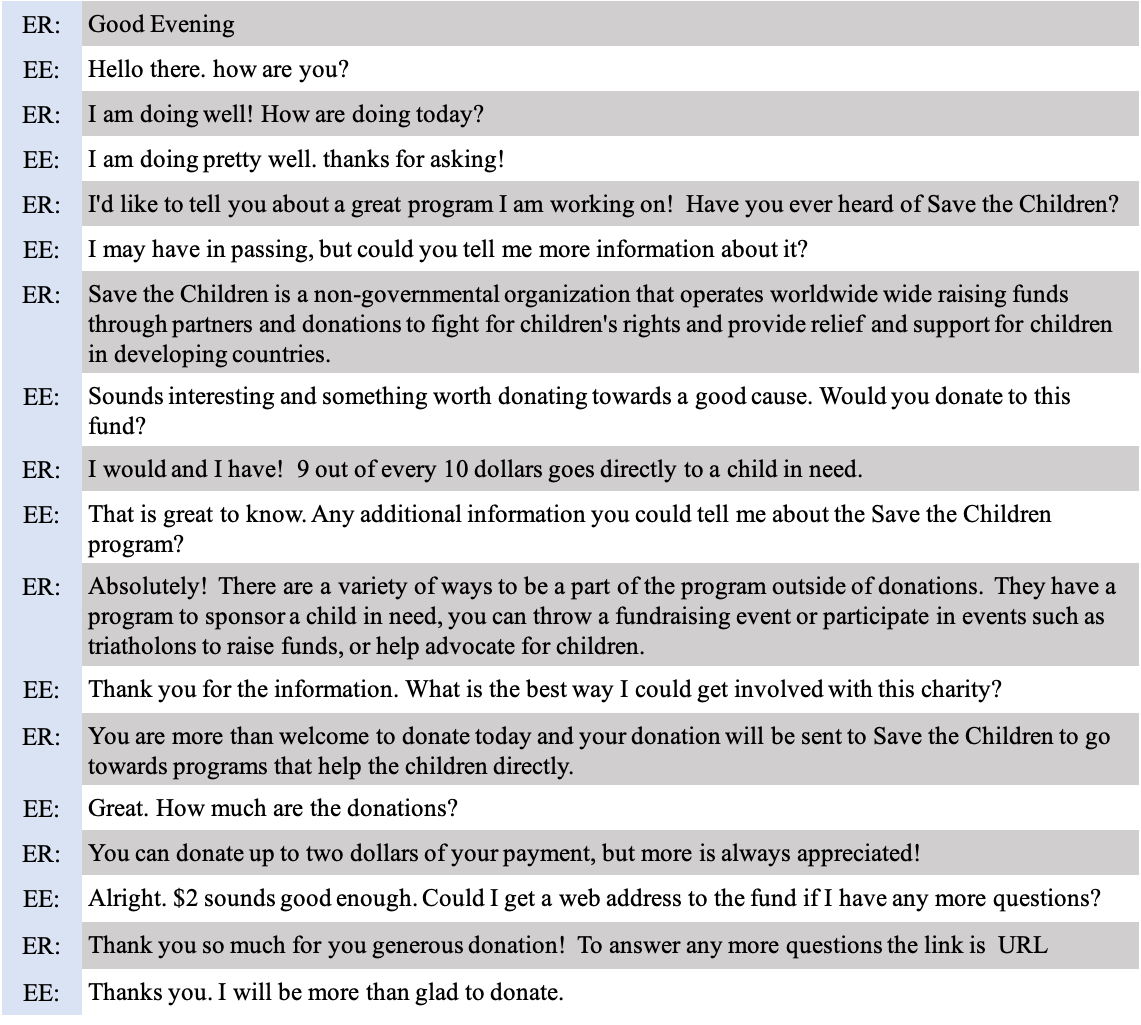}
\caption{An example (ID: $20180904\-154250\_98\_live$, donation: \$2.0, OCEAN values: 3, 3.2, 3, 3.6, 3) persuasive dialogue between persuader (ER) and persuadee (EE) from PersuasionForGood dataset. Dynamic modeling of the dialogue, utterances of ER as actions (grey), utterance of EE as states (white).} \label{example}
\vspace{-10pt}
\end{figure}

To verify our method, we use the $PersuasionForGood$~\footnote{https://convokit.cornell.edu/documentation/persuasionforgood.html} dataset of human-human dialogues. This dataset aims to facilitate the development of intelligent persuasive conversational agents and focuses on altering users' opinions and actions regarding donations to a specific charity. Through the online persuasion task, one participant as $persuader$ (ER) was asked to persuade the other as $persuadee$ (EE) to donate to a charity, Save the Children~\footnote{https://www.savethechildren.org/}. This large dataset comprises 1,017 dialogues, including annotated emerging persuasion strategies in a subset. Notably, the recorded participants' demographic backgrounds, such as the Big-Five personality traits (OCEAN), offer an opportunity to estimate the LPDs of users ($persuadees$). Additionally, each dialogue includes donation records for both $persuader$ and $persuadee$. Our focus in this work is primarily on the donation behavior of the $persuadee$, with 545 (54\%) recorded as donors and 472 (46\%) as non-donors. The dialogue data utilized in this study is represented by BERT embeddings extracted by a pre-trained BERT model~\cite{devlin2018bert} from Hugging Face\footnote{https://huggingface.co/bert-base-uncased}. Each natural language utterance of EE or ER in the dialogue is represented by a 768-dimensional BERT feature vector. The OCEAN are the real values sourced from the original dataset. For instance, in Fig.~\ref{example}, the OCEAN of a specific $persuadee$ (ID=$180904154250\_98\_live$) is presented as a 5-dimensional vector.
\vspace{-15pt}
\subsection{Experimental setup}
\begin{wraptable}[]{r}{0.65\textwidth}
\vspace{-30pt}
    \centering
    \caption{The hyperparameters across various models.}
    \begin{tabular}{|c|c|c|c|c|}
        \hline
        Model & hidden units & batch size & lr & epochs \\
        \hline
        DPPR & 1024 & 64 & 0.0001 & 100 \\
        BiCoGAN & 100 & 100 & 0.0001 & 10 \\
        RM & 256 & 64 &0.0001 & 1,000 \\
        D3QN & 256 & 60 &0.001 & 20 \\
        \hline
    \end{tabular}
    \label{hyper}
     \vspace{-15pt}
\end{wraptable}

All implementations were executed using PyTorch and models were trained on a GeForce RTX 3080 GPU (10G). Our method is a teamwork of three models, which are trained separately by using different formats of the input data. 
Hyperparameters shared by the models are listed in Table~\ref{hyper}. A five-fold cross-validation is employed to ensure the reliability of the DPPR model, which allows us to assess its generalizability to unseen data. For BiCoGAN, we set the dimension of the noise item to be the same as BERT's embedding, which is 768. In the end, we generated a total of 100 counterfactual data
by using different sets of counterfactual actions. 
To guarantee the robustness of the reward model training, dialogues featuring donation amounts surpassing \$20.0 were intentionally removed. The aim behind removing these dialogues is to reduce potential bias and ensure a more balanced training for the model. The final number of dialogues is 997, featuring 25 exchanges of utterances between EE and ER, alternating between the two roles in each dialogue. We split the data into training and testing sets using a 80/20 split. All the model is optimized using Adam~\cite{kingma2014adam} with a learning rate of 0.0001. In addition, we set the discount factor $\gamma$ in the D3QN to 0.9.

\subsection{Results} \label{results}
To facilitate the creation of the counterfactual world data~$\Tilde{D}$, we introduce the counterfactual actions~\{$a^{'}_{t}$\} as inputs into the trained BiCoGAN for~$\Tilde{D}$ generation. An initial step in this process involves establishing the counterfactual actions set \{$a^{'}_{t}$\}, which is derived through random selection from real-world action set \{$a_{t}$\}. 
However, when randomly selected as counterfactual actions, some greeting utterances will appear in the middle or end of a conversation, which is inappropriate.

\begin{figure}
\centering
\vspace{-15pt}
\includegraphics[width=\textwidth]{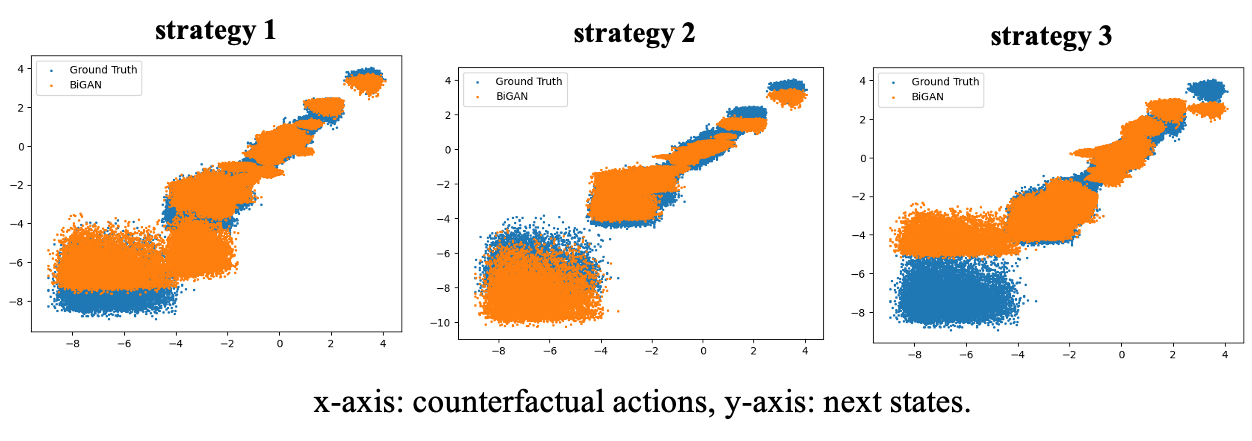}
\caption{The relationship between the counterfactual action $a^{'}_{t}$ and the next state: counterfactual case $s^{'}_{t+1}$ generated by BiCoGAN or ground truth $s_{t+1}$} \label{bicogan_situations}
\vspace{-15pt}
\end{figure}

To address this issue, we define three strategies of counterfactual action selection based on whether or not the greetings are selected: In strategy 1, all the greetings are selected, we shuffle all the real-world actions as $\{a^{'}_{t}\}$; in strategy 2, the first greeting is not selected, we remove the first utterance of persuader $\{a_{0}$\} per dialogue, then sample from the rest actions $\{a_{t}$\} - $\{a_{0}$\} as $\{a^{'}_{t}\}$; in strategy 3, all the greetings are not selected, we remove the first three utterances $\{a_{0}, a_{1}, a_{2}$\} per dialogue, then sample from the rest actions $\{a_{t}$\} - $\{a_{0}, a_{1}, a_{2}$\} as $\{a^{'}_{t}\}$. To leverage the quality of the counterfactual data across the three strategies of counterfactual action sets, we conducted a comparison between the generated counterfactual states by the trained BiCoGAN generator and the ground truth states, seen in Fig.~\ref{bicogan_situations}. We can observe that strategy 2 yields superior results as it better aligns the real world's next state, $s_{t+1}$, and counterfactual next state, $s^{'}_{t+1}$, compared to the other two strategies. Finally, counterfactual data~$\tilde{D}$ can be represented as 
\begin{equation}
\begin{aligned}
\tilde{D} &= \{\tilde{D}_{0}, \tilde{D}_{1}, ..., \tilde{D}_{i}, ..., \tilde{D}_{N-1}\} \\
&= \{(s^{'}_{0}, a^{'}_{0}, s^{'}_{1}, a^{'}_{1}, ..., s^{'}_{t}, a^{'}_{t}, ..., s^{'}_{T-1})^{\Sigma-1}_{j=0}\}^{N-1}_{i=0},
\end{aligned}
\end{equation}
where $T$ typically represents the total number of steps or time points in the sequence, $\Sigma$ is the number of dialogues, and $N$ is the number of counterfactual databases.

\begin{figure}[!htb]
\centering
   \begin{minipage}{0.48\textwidth}
     \centering
     \includegraphics[width=\linewidth]{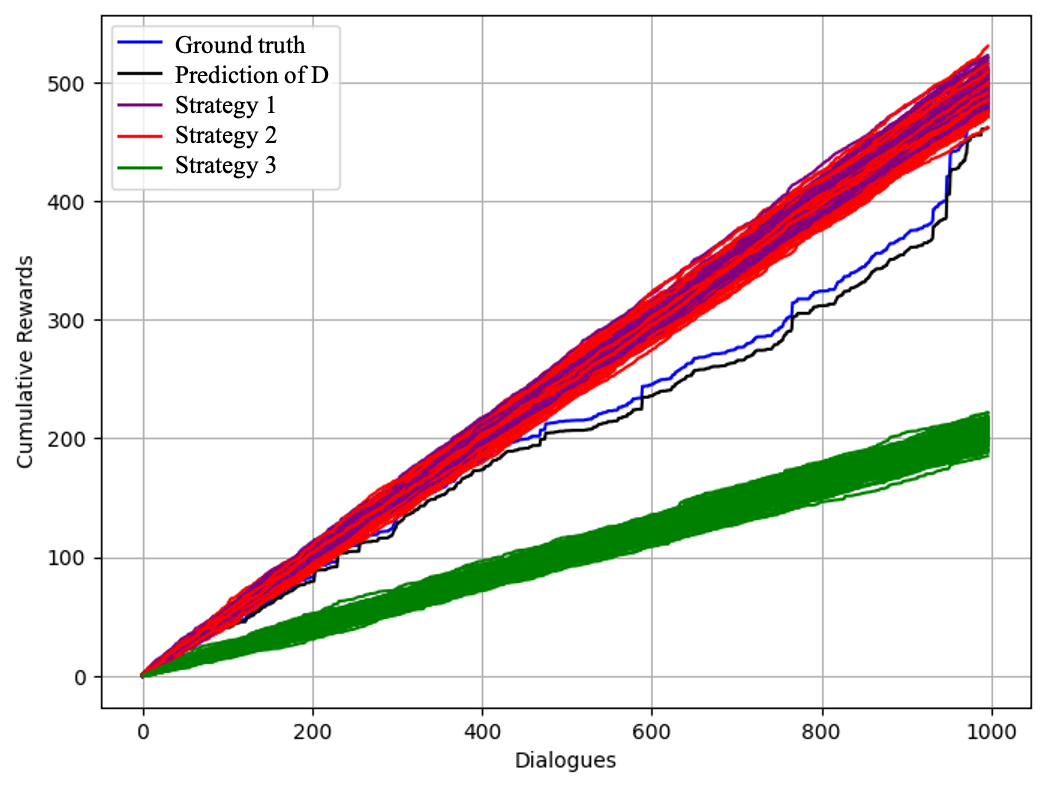}
     \caption{The (BiCoGAN + DPPR) reward predictions of counterfactual data on three strategies and real data $D$ (black), compared with ground-truth (blue).}
     \label{reward_p}
   \end{minipage}\hfill
   \begin{minipage}{0.48\textwidth}
     \centering
     \includegraphics[width=\linewidth]{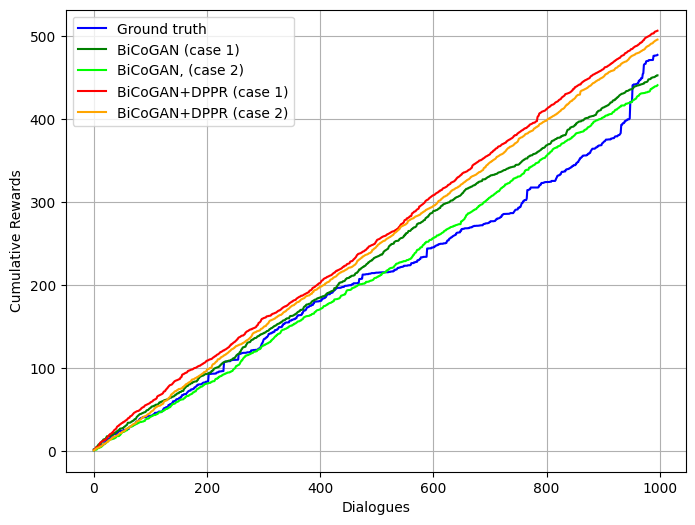}
     \caption{Comparison of cumulative rewards in dialogues between ground truth and counterfactual cases utilizing BiCoGAN or BiCoGAN+DPPR.}
     \label{cumulative_reward}
   \end{minipage}
   \vspace{-15pt}
\end{figure}

Following the estimation of the dynamics model and the creation of an augmented dataset~$\Tilde{D}$ through counterfactual reasoning, the subsequent step involves training policies on the counterfactual data~$\Tilde{D}$ to optimize the Q values and improve the future predicted cumulative rewards. 
To ensure fairness in policy learning, we aim for balanced predicted cumulative rewards of~$\Tilde{D}$, with 50 counterfactual databases exceeding and 50 counterfactual databases falling short of the ground truth, thus forming our final counterfactual data$~\Tilde{D}$. (cf: Section~\ref{counter_data}). 
For cumulative reward computation, we first input the dialogue into the reward model (Equation (\ref{reward_fun})), then perform a cumulative sum operation on the predicted reward of the trained reward model. Fig.~\ref{reward_p} shows the cumulative reward prediction for the three strategies, where strategies 1 and 2 are mostly higher than the ground truth's, while strategy 3 is lower than the ground truth. 

\begin{figure}[h]
\centering
\includegraphics[width=0.9\textwidth]{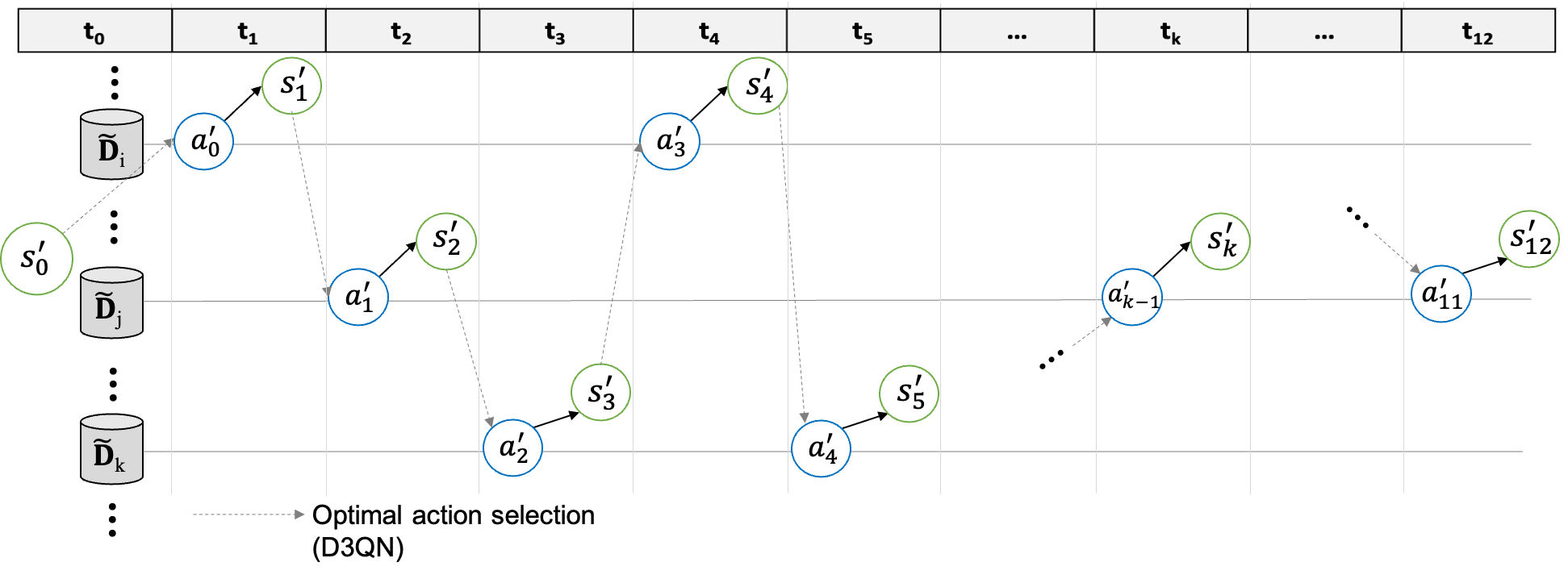}
\caption{The process of counterfactual dialogue produced for reward prediction during the policy learning.} \label{newdialogue}
\end{figure}

During policy learning with the D3QN model, the process involves computing Q values for 100 action candidates from 100 counterfactual databases at each state. The model selects the action with maximum Q value for each time step. As a result, this process generates a counterfactual dialogue and then utilizes the trained reward model (cf: Equation (\ref{reward_fun})) to predict the reward. Determining whether the counterfactual dialogue in natural language signifies a donation or not will be our focus for future work. For the optimization of D3QN, we established two distinct cases based on the optimization time for the loss function. In \textbf{Case 1}, where the loss function is optimized once per dialogue. For example, the sequence of $j$-th dialogue starts from state $s^{'}_{j, 0}$. It proceeds with a series of action-state pairs $a^{'}_{j, 0}$, $s^{'}_{j, 1}$, ..., $a^{'}_{j, i}$, $s^{'}_{j, i}$, ..., $a^{'}_{j, 12}$, $s^{'}_{j,12}$, for each state $s^{'}_{j, i}$, based on $k = argmax(Q(s^{'}_{j, i}, a^{'}_{j})^{N}_{i=0})$ ($N$=99) to select the next action $a^{'}_{j, k}$ from the $j$-th dialogue of $\Tilde{D}_{k}$. In contrast, \textbf{Case 2} differs as the loss function optimization takes place once per state. The action selection process is illustrated in Fig.~\ref{newdialogue}

\begin{figure}[h]
\centering
\includegraphics[width=0.9\textwidth]{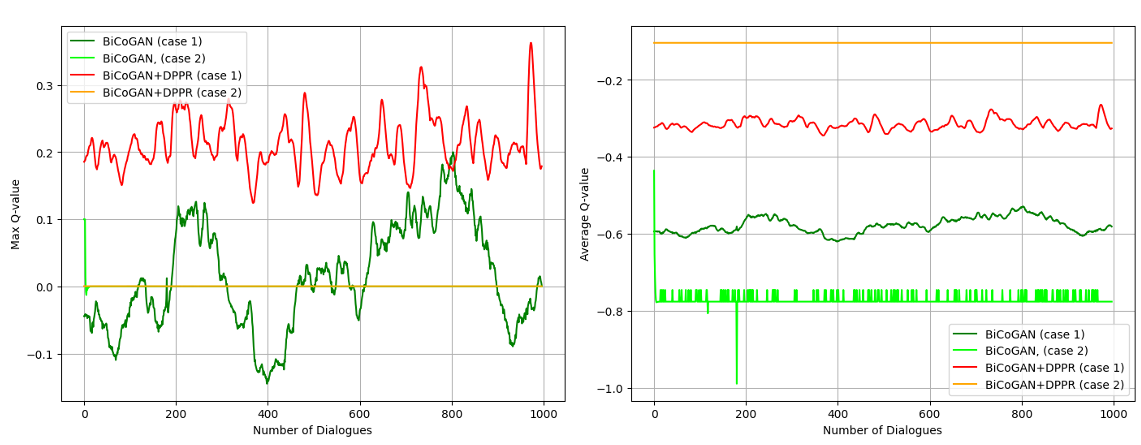}
\caption{Comparison of maximum and average Q values in dialogues between ground truth and counterfactual cases utilizing BiCoGAN or BiCoGAN+DPPR.} \label{fig:qvalue}
\end{figure}
\vspace{-3pt}
After the policy learning, we obtain a corresponding counterfactual dialogue for donation prediction, aligning with each real-world dialogue. The sequences of real-world dialogues begin from $s_{0}$, while the counterfactual dialogues starts from $s^{'}_{0}$, where maintaining the equivalence of $s_{0}$ = $s^{'}_{0}$. In Fig.~\ref{cumulative_reward}, the cumulative rewards of both BiCoGAN and BiCoGAN+DPPR models in two cases are higher than the ground truth overall. 
It can be observed that the BiCoGAN+DPPR (case 1) gets the best cumulative rewards and BiCoGAN+DPPR (case 2) obtains the second results over dialogues. The final cumulative rewards/donation amounts are \$506.82 (+\$29.2) and \$496.15 (+\$18.53) respectively, higher than ground truth cumulative rewards \$477.62 and the BiCoGAN in case 1 and case 2, achieved \$453.02 (-\$24.6) and \$441.11 (-\$36.51), respectively. 

Furthermore, when estimating the Q-values of the learned optimal policy (Fig.~\ref{fig:qvalue}), the optimal policy derived from BiCoGAN with LPDs exhibits higher estimated Q-values, both in maximum and average, compared to the single BiCoGAN model. In Case 1 of BiCoGAN+DPPR, the maximum Q-values surpass Case 2, while the average Q-values are lower in Case 1 than in Case 2. This discrepancy arises from weight optimization in D3QN primarily at the dialogue's end, potentially leading to actions with exceptionally high Q-values.
\vspace{-5pt}
\subsection{Ablation Study}
\subsubsection{Impact of window size} 
\begin{figure}
\centering
\vspace{-10pt}
\includegraphics[width=0.95\textwidth]{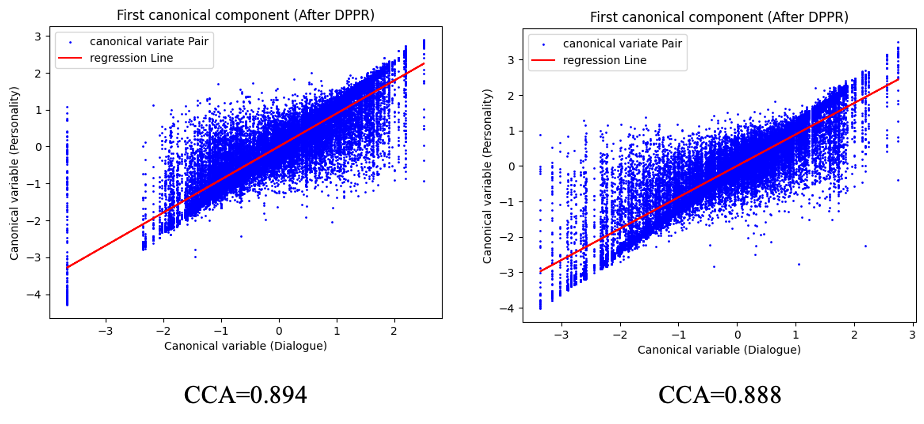}
\caption{The canonical coefficient of the top two CCA components.} \label{fig:cca}
\vspace{-25pt}
\end{figure}

plays an essential role in the DPPR model, it will influence the accuracy of prediction and determine how much information is meaningful for the individual LPDs estimation. We separately set the window size as 1, 2, 3, and 4 turns, each turn includes exchange utterances between 
\begin{wraptable}{r}{0.5\textwidth}
    \centering
    \vspace{-27pt}
    \caption{Performance of DPPR model with varied window sizes.}
    \begin{tabular}{|c|c|c|c|c|c|}
        \hline
        Win size & MSE & RMSE & MAPE & R2 &MAE \\
        \hline
        1 turn  &0.166 &0.407 &0.092 &0.830 &0.254 \\ 
        2 turns &0.258 &0.508 &0.119 &0.641 &0.323 \\
        4 turns &0.441 &0.664 &0.178 &0.387 &0.474 \\
        8 turns &0.488 &0.698 &0.195 &0.319 &0.518 \\
        \hline
    \end{tabular}
    \label{tab:dppr}
    \vspace{-25pt}
\end{wraptable} a $persuader$ and a $persuadee$. The evaluation\footnote{MSE - Mean Squared Error; RMSE - Root Mean Squared Error; MAPE - Mean Absolute Percentage Error; R2 - R-squared (Coefficient of Determination); MAE - Mean Absolute Error} results of DPPR on different window size is shown in Table~\ref{tab:dppr}, we can observe that when the window sizes decrease, the precision increase, so we choose one-turn to train DPPR model for counterfactual data building. To further investigate the one-turn trained model, we leverage the first top two canonical components, shown in Fig.~\ref{fig:cca}, the CCA values for each are 0.894 and 0.888, respectively, which indicate there exist relatively high correlations between the one-turn utterance and personality.
\vspace{-5pt}
\section{Conclusion}\label{conclusion}
In this paper, we highlight the limitations of existing persuasive conversation systems, particularly in adapting to individual user traits. 
We augmented the BiCoGAN model, which creates counterfactual data that simulates an alternative world, with our DPPR models to enable awareness of latent user dimensions during persuasive interactions. This facilitated an optimized conversation flow using the D3QN model. In the 
experiments conducted on the PersuasionForGood dataset, our approach showcased 
superiority over existing method, BiCoGAN, and ground truth, demonstrating the efficacy of leveraging latent traits to enhance persuasion outcomes.
\bibliographystyle{splncs04}
\bibliography{mybibliography}
\end{document}